# Fronthaul-Aware Group Sparse Precoding and Signal Splitting in SWIPT C-RAN

Yanjie Dong, *Student Member, IEEE*[‡], Md. Jahangir Hossain, *Member, IEEE*[†],
Julian Cheng, *Senior Member, IEEE*[†], and Victor C. M. Leung, *Fellow, IEEE*[‡]

[‡]Department of Electrical and Computer Engineering, The University of British Columbia, Vancouver, BC, Canada
[†]School of Engineering, The University of British Columbia, Kelowna, BC, Canada
Emails: {ydong16, vleung}@ece.ubc.ca, {julian.cheng, jahangir.hossain}@ubc.ca

*Abstract*—We investigate the precoding, remote radio head (RRH) selection and signal splitting in the simultaneous wireless information and power transferring (SWIPT) cloud radio access networks (C-RANs). The objective is to minimize the power consumption of the SWIPT C-RAN. Different from the existing literature, we consider the nonlinear fronthaul power consumption and the multiple antenna RRHs. By switching off the unnecessary RRHs, the group sparsity of the precoding coefficients is introduced, which indicates that the precoding process and the RRH selection are coupled. In order to overcome these issues, a group sparse precoding and signal splitting algorithm is proposed based on the majorization-minimization framework, and the convergence behavior is established. Numerical results are used to verify our proposed studies.

## I. INTRODUCTION

Recently, it has been demonstrated that the 1 and 11 Mbps wireless transmissions consume, respectively, 14.5 and 59.2 $\mu$W in the passive Wi-Fi [1]. Therefore, with the development of the energy harvesting (EH) chipsets (e.g., Powercastco[1]), the radio frequency (RF) energy harvesting technology becomes an attractive solution to provide ubiquitous low power communications with a sustainable energy supply [2]. As an important use case of the RF-EH technology, the simultaneous wireless information and power transferring (SWIPT) technique has acquired recent attention due to the capability of feeding the receivers with information and energy simultaneously [3]. However, the current circuits do not allow a receiver to detect information and scavenge energy from a single signal stream. This fact leads to several research efforts in the design of reception schemes for the systems with SWIPT capability, and two practical receiver schemes are proposed, namely, the time switching (TS) scheme and the signal splitting (SS) scheme. It has been shown that the SS scheme is a general case of the TS scheme [3]; therefore, we investigate the SS scheme based SWIPT technique in this work.

Meanwhile, cloud radio access network (C-RAN) has become one of the candidate architectures for the 5-th generation (5G) wireless networks due to its cost effectiveness and deployment flexibility [4]. A typical C-RAN consists of a baseband unit (BBU) pool and a set of low-power and low-cost remote radio heads (RRHs). All the RRHs are connected to the BBU pool via optical fiber links. The BBU pool performs the processor-heavy ciphering functions, and the RRHs perform basic radio frequency processing functions [5]. As a result, a significant performance improvement can be achieved via joint scheduling and information processing such as coordinated transmission [6]. It has also been reported that the coordinated transmission can transmit significantly more energy to the energy receivers compared with the single-antenna transmitter [7]. Due to an efficient interference mitigation scheme, the RRHs can be ultra-densely deployed in the coverage area of the C-RAN such that the distance between transceivers is reduced. Hence, the spectrum efficiency is enhanced.

In order to avoid the energy depletion of the lower power devices, the SWIPT technique will become a key component of the C-RAN architecture. The integration of the SWIPT technique into the C-RAN also introduces many interesting and new challenges to the resource allocation, which needs to be solved carefully to bridge the gap between theory and practice [5]. For example, the power consumption of the SWIPT C-RAN consists of the transmission power cost, the circuit power cost and the fronthaul power cost. To minimize the transmission power cost, the precoding vector should be carefully designed. To reduce the circuit power cost and the fronthaul power cost, the RRH selection mechanism should be considered. As a result, a joint precoding, RRH selection and signal splitting algorithm is required in the SWIPT C-RAN architecture.

### A. Related Work and Contributions

The SWIPT technique finds the applications in various wireless networks, such as point-to-point networks [8], point-to-multipoint networks [9], [10], multipoint-to-multipoint networks [11]–[14] and wireless cooperative networks [15]–[18]. In particular, the resource allocation in the multipoint-to-multipoint networks with SWIPT capability has focused on the power minimization [11], energy efficiency [12], the joint uplink-and-downlink design [13] and the robust design [14]. However, the impact of capacity-limited fronthaul links, which may cause a mismatch between the resource allocation and the fronthaul capacity, is ignored in previous works [11]–[14]. Assuming that each receiver shares a capacity-limited backhaul[2] link with a fixed data rate, the authors in [19] developed a beamforming algorithm to minimize the power consumption of the distributed antenna systems while constrains the received signal-to-interference-plus-noise ratio (SINR) of the eavesdropper. However, the assumption made in [19] can cause some backhaul links under utilization when the acquired data rate of the associated information receivers is lower than the obtained fixed data from the backhaul links.

---

[1]http://www.powercastco.com/products/powerharvester-receivers/

[2]The backhaul in the distributed antenna systems has the similar function with the fronthaul in the C-RAN architecture.

On the other hand, the objective of the proposed algorithms in [20] is to minimize the power consumption of the C-RANs without SWIPT capability via joint beamforming and RRH selection. The power consumption in [20] is the summation of the transmission power, the circuit power and the fronthaul links power consumption. A linear model is used in [20] to obtain the fronthaul links power consumption, which ignores the nonlinearity of the media of the fronthaul links. Besides, each RRH of the C-RAN is assumed to be equipped with a single antenna in [20]. As a result, the proposed beamforming and RRH selection algorithm in [20] cannot be used for the multiple antenna RRHs in the SWIPT C-RANs.

Motivated by the previous works [19], [20], we investigate the power consumption minimization problem by jointly considering the precoding, the RRH selection and the signal splitting in the SWIPT C-RANs. In particular, the RRHs coordinately deliver the wireless signals to the information detection module (IDM) and energy harvesting module (EHM) of each user equipment (UE). Different from [20], the fronthaul links power consumption increases nonlinearly with the data rate on the fronthaul links, where the slope of the power cost-data rate curve increases with the increasing data rate. Considering the RRH selection, an RRH is allowed to be switched off when all the precoding coefficients associated with the RRH are set to zeros, which refers to group sparsity for the precoding coefficients [21]. Hence, the precoding process and the RRH selection are tightly coupled. Inspired by [22], we leverage the majorization-minimization (MM) framework to decouple the precoding and the RRH selection. As a result, we propose a group sparse precoding and signal splitting (GsPSs) algorithm to minimize the power consumption, and study the convergence of the proposed algorithm.

### B. Organization and Notations

The remainder of this paper is organized as follows. Section II introduces the system model and problem formulation. The feasibility of the formulated problem, the proposed GsPSs algorithm and the convergence are discussed in Section III. Numerical results are presented in Section IV, and conclusions are drawn in Section V.

*Notations:* Vectors and matrices are shown in bold lowercase letters and bold uppercase letters, respectively. $\mathbb{C}$ denotes the set of complex values. $\|\cdot\|_\mathrm{F}$ and $\|\cdot\|_0$ refer to Frobenius norm and $\ell_0$-norm, respectively. $\sim$ stands for "distributed as". $\boldsymbol{I}_N$ and $\boldsymbol{0}_{N\times M}$ denote, respectively, an $N$ dimensional identity matrix and a zero matrix with $N$ rows and $M$ columns. The expectation of a random variable is denoted as $\mathbb{E}[\cdot]$. $\mathrm{vec}[\boldsymbol{W}]$ obtains a vector by stacking the columns of $\boldsymbol{W}$ under the other. $\{\boldsymbol{w}_n\}_{n\in\mathcal{N}}$ represents the set made of $\boldsymbol{w}_n$, $n\in\mathcal{N}$. For a square matrix $\boldsymbol{W}$, $\boldsymbol{W}^\mathrm{H}$ and $\mathrm{Tr}(\boldsymbol{W})$ denote its conjugate transpose and trace, respectively. $\boldsymbol{W}\succeq\boldsymbol{0}$ and $\boldsymbol{W}\succ\boldsymbol{0}$ denote that $\boldsymbol{W}$ is a positive semidefinite and a positive definite matrix.

## II. SYSTEM MODEL AND PROBLEM FORMULATION

### A. Overall Description

We consider the downlink transmission scenario of the SWIPT C-RAN, which consists of a BBU pool, a set of RRHs and a set of UEs. Let $\mathcal{M}=\{1,2,\ldots,M\}$ and $\mathcal{N}=\{1,2,\ldots,N\}$ denote the set of RRHs and UEs, respectively. The $m$-th RRH is equipped with $i_m$ antennas, and each UE is equipped with a single antenna. As a result, the access link from all RRHs to each UE is a MISO channel with the number of transmission antennas as $N_T=\sum_{m=1}^M i_m$. We assume that SWIPT C-RAN operates in the time division duplex mode such that each RRH can obtain the channel state information (CSI) perfectly via exploiting the uplink reciprocity and reports the obtained CSI to the BBU pool via the common public radio interface. Each RRH connects to the BBU pool via a capacity-limited optical fiber fronthaul link. We consider the frame-based frequency-nonselective fading channels with unit duration for each frame; therefore, the words "power" and "energy" can be used interchangeably. Via a signal splitter, each UE can split the incoming signal stream into two substreams for the IDM and the EHM. Specifically, the power splitting ratio for the IDM and EHM are, respectively, denoted by $\rho_n^2 \geq 0$ and $\varrho_n^2 \geq 0$ with $\rho_n^2 + \varrho_n^2 \leq 1$.

At the BBU pool, the information-bearing signal $s_n$ for the $n$-th UE is processed by the central encoder via two procedures: precoding and compression. Generally, the precoding procedure controls the multi-user interference, and the precoded signal vector for the $m$-th BBU-RRH fronthaul link is denoted by $\tilde{\boldsymbol{x}}_m = \sum_{n=1}^N \boldsymbol{w}_{m,n}s_n$ where $\mathbb{E}[|s_n|^2]=1$; $\boldsymbol{w}_{m,n}\in\mathbb{C}^{i_m\times 1}$ is the precoding vector used for the transmission from the $m$-th RRH to the $n$-th UE. Each precoded signal vector $\tilde{\boldsymbol{x}}_m$ is compressed in order to be delivered over the $m$-th BBU-RRH fronthaul link. The compressed signal vector for the $m$-th BBU-RRH fronthaul link is denoted by $\boldsymbol{x}_m = \sum_{n=1}^N \boldsymbol{w}_{m,n}s_n + \boldsymbol{q}_m$, where $\boldsymbol{q}_m \in \mathbb{C}^{i_m\times 1}$ denotes the quantization noise and is modeled as a circularly-symmetric complex Gaussian random vector $\mathcal{CN}(\boldsymbol{0}_{i_m\times 1}, \boldsymbol{Q}_m)$ with covariance matrix $\boldsymbol{Q}_m \triangleq \mathbb{E}[\boldsymbol{q}_m\boldsymbol{q}_m^\mathrm{H}]$. In this work, we assume that the encoder can compress the precoded signal for each antenna independently. In other words, the covariance matrix $\boldsymbol{Q}_m$ is a diagonal matrix, $m\in\mathcal{M}$.

By defining $\boldsymbol{w}_n \triangleq \mathrm{vec}([\boldsymbol{w}_{1,n},\boldsymbol{w}_{2,n},\ldots,\boldsymbol{w}_{M,n}])$ and $\boldsymbol{B}_m \triangleq \left[\boldsymbol{0}_{i_m\times\sum_{k=1}^{m-1}i_k}, \boldsymbol{I}_{i_m}, \boldsymbol{0}_{i_m\times\sum_{k=m+1}^M i_k}\right]$, the compact form of the compressed signal vector for the $m$-th BBU-RRH fronthaul link is denoted as

$$\boldsymbol{x}_m = \sum_{n=1}^N \boldsymbol{B}_m \boldsymbol{w}_n s_n + \boldsymbol{q}_m \quad (1)$$

with covariance matrix $\mathbb{E}[\boldsymbol{x}_m\boldsymbol{x}_m^\mathrm{H}] = \sum_{n=1}^N \boldsymbol{B}_m \boldsymbol{w}_n \boldsymbol{w}_n^\mathrm{H} \boldsymbol{B}_m^\mathrm{H} + \boldsymbol{\Lambda}_m^2$, where $\boldsymbol{\Lambda}_m = \boldsymbol{Q}_m^{\frac{1}{2}}$, $m\in\mathcal{M}$.

According to the rate-distortion theory, there exists at least one compression codebook under a Gaussian test channel such that the achievable information transmission rate of the $m$-th BBU-RRH fronthaul link is denoted by [23]

$$g_m\left(\{\boldsymbol{w}_n\}_{n\in\mathcal{N}}, \boldsymbol{\Lambda}_m\right) \quad (2)$$
$$= \log\det\left(\sum_{n=1}^N \boldsymbol{B}_m \boldsymbol{w}_n \boldsymbol{w}_n^\mathrm{H} \boldsymbol{B}_m^\mathrm{H} + \boldsymbol{\Lambda}_m^2\right) - 2\log\det(\boldsymbol{\Lambda}_m).$$

where $\boldsymbol{\Lambda}_m \succ \boldsymbol{0}$ when the $m$-th RRH is on.

The channel coefficient vector from the $m$-th RRH to the $n$-th UE is denoted as $\boldsymbol{h}_{m,n} \sim \mathcal{CN}\left(\boldsymbol{0}_{i_m \times 1}, d_{m,n}^{-\chi}\boldsymbol{I}_{i_m}\right)$ [24, and references therein]. Here, $d_{m,n}$ and $\chi$ are, respectively, the distance between the $m$-th RRH and the $n$-th UE, and the pathloss exponent. By defining $\boldsymbol{h}_n \triangleq \text{vec}\left([\boldsymbol{h}_{1,n}, \boldsymbol{h}_{2,n}, \ldots, \boldsymbol{h}_{M,n}]\right)$, the received baseband signal of the $n$-th UE in the compact form is obtained as

$$\tilde{y}_n = \sum_{k=1}^{N} \boldsymbol{h}_n^{\text{H}}\boldsymbol{w}_k s_k + \sum_{m=1}^{M} \boldsymbol{h}_n^{\text{H}}\boldsymbol{q}_m + z_{n,\text{c}} \quad (3)$$

where $z_{n,\text{c}} \sim \mathcal{CN}\left(0, \sigma_{n,\text{c}}^2\right)$ is the additive white Gaussian noise (AWGN) at the $n$-th UE.

As a result, the received signal of the IDM and the received SINR of the $n$-th UE are, respectively, expressed as

$$y_n = \rho_n \left( \sum_{k=1}^{N} \boldsymbol{h}_n^{\text{H}}\boldsymbol{w}_k s_k + \sum_{m=1}^{M} \boldsymbol{h}_n^{\text{H}}\boldsymbol{q}_m + z_{n,\text{c}} \right) + z_{n,\text{p}} \quad (4)$$

and

$$\text{SINR}_n = \frac{\left|\boldsymbol{h}_n^{\text{H}}\boldsymbol{w}_n\right|^2}{\sum_{k=1,k\neq n}^{N} \left|\boldsymbol{h}_n^{\text{H}}\boldsymbol{w}_k\right|^2 + P_n^{\text{Compress}} + \sigma_{n,\text{c}}^2 + \frac{\sigma_{n,\text{p}}^2}{\rho_n^2}} \quad (5)$$

where $P_n^{\text{Compress}} \triangleq \sum_{m=1}^{M} \left\|\boldsymbol{h}_n^{\text{H}}\boldsymbol{B}_m^{\text{H}}\boldsymbol{\Lambda}_m\right\|_{\text{F}}^2$; $z_{n,\text{p}} \sim \mathcal{CN}\left(0, \sigma_{n,\text{p}}^2\right)$ denotes the additional signal processing noise during the baseband conversion [8]–[19].

On the other hand, the input signal power at the EHM of the $n$-th UE is given by

$$P_n^{\text{Input}} = \varrho_n^2 \left( \sum_{k=1}^{N} \left|\boldsymbol{h}_n^{\text{H}}\boldsymbol{w}_k\right|^2 + P_n^{\text{Compress}} + \sigma_{n,\text{c}}^2 \right). \quad (6)$$

*B. Power Consumption Model*

When a specific RRH is on, a certain amount of circuit power is consumed for signal processing [25]. Some circuit power consumption is still required for fast wake up for the RRH in sleep mode. Hence, the power consumption of the $m$-th RRH is denoted by

$$P_m^{\text{RRH}} = \begin{cases} \frac{1}{\phi_m} \left( \sum_{n=1}^{N} \|\boldsymbol{B}_m\boldsymbol{w}_n\|_{\text{F}}^2 + \|\boldsymbol{\Lambda}_m\|_{\text{F}}^2 \right) \\ \qquad + P_m^{\text{Active}}, \sum_{n=1}^{N} \|\boldsymbol{B}_m\boldsymbol{w}_n\|_{\text{F}}^2 \neq 0 \\ P_m^{\text{Sleep}}, \text{otherwise} \end{cases} \quad (7)$$

where the constant $\phi_m$ denotes the power amplifier efficiency of the $m$-th RRH; the circuit power consumption of an active RRH is given as $P_m^{\text{Active}} \triangleq P_m^{\text{SP}}\left(0.87 + 0.1 i_m + 0.03 i_m^2\right)$ with the signal processing power consumption $P_m^{\text{SP}}$ [25]; $P_m^{\text{Sleep}}$ is the circuit power consumption in sleep mode.

Different from [20], we consider the nonlinear power consumption of the $m$-th BBU-RRH fronthaul link as

$$P_m^{\text{BH}} = \zeta_m g_m^2\left(\{\boldsymbol{w}_n\}_{n \in \mathcal{N}}, \boldsymbol{\Lambda}_m\right) \quad (8)$$

where $\zeta_m$ can be obtained via curve fitting techniques. The nonlinear model of the power consumption can be justified as follows. The higher data rate via the fronthaul fiber links results in a higher transmission voltage of the signals under the same bit error rate. Therefore, according to the Ohm's law, more power is consumed on the same fronthaul fiber links with unchanged resistance[3].

*C. Problem Formulation*

Our objective is to minimize the power consumption of the SWIPT C-RAN while guarantee the data rate and the delivered power for the UEs via joint precoding, RRH selection and signal splitting. Considering the power consumption on transmission, circuit and fronthaul links, we formulate the power consumption optimization problem as

$$\min_{\substack{\{\boldsymbol{w}_n, \rho_n, \varrho_n\}_{n \in \mathcal{N}} \\ \{\boldsymbol{\Lambda}_m\}_{m \in \mathcal{M}}}} P^{\text{Total}}\left(\{\boldsymbol{w}_n\}_{n \in \mathcal{N}}, \{\boldsymbol{\Lambda}_m\}_{m \in \mathcal{M}}\right) \quad (9a)$$

$$\text{s.t.} \sum_{n=1}^{N} \|\boldsymbol{B}_m\boldsymbol{w}_n\|_{\text{F}}^2 + \|\boldsymbol{\Lambda}_m\|_{\text{F}}^2 \leq P_m^{\max}, \forall m \quad (9b)$$

$$\text{SINR}_n \geq \Gamma_n^{\text{req}}, \forall n \quad (9c)$$

$$P_n^{\text{Input}} \geq P_n^{\text{req}}, \forall n \quad (9d)$$

$$\rho_n^2 + \varrho_n^2 \leq 1, \forall n \quad (9e)$$

where the power consumption $P^{\text{Total}}\left(\{\boldsymbol{w}_n\}_{n \in \mathcal{N}}, \{\boldsymbol{\Lambda}_m\}_{m \in \mathcal{M}}\right)$ is defined in (10); $P_m^{\max}$ in constraints (9b) denotes the maximum transmission power of the $m$-th RRH; $\Gamma_n^{\text{req}}$ and $P_n^{\text{req}}$ are, respectively, the data rate requirement for the IDM and the required input power for the EHM of the $n$-th UE.

*Remark 1:* With the energy harvester model in [16], harvesting the quantity of $\hat{P}_n^{\text{req}}$ power requires the input power $P_n^{\text{req}} = \frac{1}{a_n}\ln\left(\frac{M_n + \hat{P}_n^{\text{req}}\exp(a_n b_n)}{M_n - \hat{P}_n^{\text{req}}}\right)$, where $M_n$, $a_n$ and $b_n$ are the shaping parameters of the nonlinear energy harvester.

## III. JOINT RRH SELECTION AND BEAMFORMER DESIGN

We observe that the optimization problem (9) is non-convex due to the non-convex feasible region and the non-convex objective function. Adding (9c) to (9d) and performing some

---

[3]In practice, several impact factors influence resistance. We assume the resistance unchanged when the same fronthaul fiber link is used.

$$P^{\text{Total}}\left(\{\boldsymbol{w}_n\}_{n \in \mathcal{N}}, \{\boldsymbol{\Lambda}_m\}_{m \in \mathcal{M}}\right) \triangleq \sum_{m=1}^{M} \frac{1}{\phi_m}\left(\sum_{n=1}^{N}\|\boldsymbol{B}_m\boldsymbol{w}_n\|_{\text{F}}^2 + \|\boldsymbol{\Lambda}_m\|_{\text{F}}^2\right) + \sum_{m=1}^{M} P_m^{\text{Active}}\left\|\sum_{n=1}^{N}\|\boldsymbol{B}_m\boldsymbol{w}_n\|_{\text{F}}^2\right\|_0$$

$$+ \sum_{m=1}^{M} P_m^{\text{Sleep}}\left(1 - \left\|\sum_{n=1}^{N}\|\boldsymbol{B}_m\boldsymbol{w}_n\|_{\text{F}}^2\right\|_0\right) + P^{\text{BBU}} + \sum_{m=1}^{M}\zeta_m g_m^2\left(\{\boldsymbol{w}_n\}_{n \in \mathcal{N}}, \boldsymbol{\Lambda}_m\right) \quad (10)$$

algebraic manipulations, we relax the feasible region of the optimization problem (9) as

$$\mathcal{F} = \{\sum_{n=1}^{N} \|\boldsymbol{B}_m \boldsymbol{w}_n\|_F^2 + \|\boldsymbol{\Lambda}_m\|_F^2 \leq P_m^{\max}, \forall m \quad (11a)$$

$$\frac{|\boldsymbol{h}_n^H \boldsymbol{w}_n|^2}{\Gamma_n^{\text{req}}} \geq \sum_{k=1, k \neq n}^{N} |\boldsymbol{h}_n^H \boldsymbol{w}_k|^2 + \frac{\sigma_{n,p}^2}{\rho_n^2}$$
$$+ P_n^{\text{Compress}} + \sigma_{n,c}^2, \forall n \quad (11b)$$

$$\sqrt{\frac{1+\Gamma_n^{\text{req}}}{\Gamma_n^{\text{req}}}} \boldsymbol{h}_n^H \boldsymbol{w}_n \geq \sqrt{\frac{\sigma_{n,p}^2}{\rho_n^2} + \frac{P_n^{\text{req}}}{\varrho_n^2}}, \forall n \quad (11c)$$

$$\rho_n^2 + \varrho_n^2 \leq 1, \forall n \quad (11d)$$

$$\text{Imag}(\boldsymbol{h}_n^H \boldsymbol{w}_n) = 0, \forall n\} \quad (11e)$$

where the operator $\text{Imag}(\cdot)$ in constraints (11e) represents the imaginary part of a complex value. As a result, the relaxation of the optimization problem (9) is denoted as

$$\min_{\{\boldsymbol{w}_n, \rho_n, \varrho_n\}_{n \in \mathcal{N}}, \{\boldsymbol{\Lambda}_m\}_{m \in \mathcal{M}}} P^{\text{Total}}\left(\{\boldsymbol{w}_n\}_{n \in \mathcal{N}}, \{\boldsymbol{\Lambda}_m\}_{m \in \mathcal{M}}\right)$$
$$\text{s.t. } (11a) - (11e). \quad (12)$$

It has been proved in [11, Proposition 3.1] that a solution to (12) is also a solution to (9) when a certain relationship of the system parameters $\sigma_{n,c}^2$, $\sigma_{n,p}^2$, $\Gamma_n^{\text{req}}$ and $P_n^{\text{req}}$ is satisfied. Without loss of generality, we assume that this relationship in [11, Proposition 3.1] is satisfied.

### A. Feasibility Analysis

The problem (12) can be infeasible under certain channel conditions when the values of $\Gamma_n^{\text{req}}$ and $P_n^{\text{req}}$ are too high. In order to check the feasibility of the region in (11), we first reformulate the region as the intersection of a set of convex cones. Introducing a set of auxiliary variables $\{\theta_n\}_{n \in \mathcal{N}}$, the constraints in (11b) can be reformulated as the intersection of a second-order cone and positive semidefinite cone as

$$\frac{\boldsymbol{h}_n^H \boldsymbol{w}_n}{\sqrt{\Gamma_n^{\text{req}}}} \geq \left\|\left[\boldsymbol{h}_n^H \boldsymbol{W}_{-k}, \boldsymbol{h}_n^H \boldsymbol{\Sigma}, \sigma_{n,c}, \theta_n\right]\right\|_F, \forall n \quad (13)$$

$$\begin{bmatrix} \theta_n & \sigma_{n,p}^{\frac{1}{2}} \\ \sigma_{n,p}^{\frac{1}{2}} & \rho_n \end{bmatrix} \succeq 0, \forall n \quad (14)$$

where $\boldsymbol{W}_{-k} \triangleq [\boldsymbol{w}_1, \ldots, \boldsymbol{w}_{k-1}, \boldsymbol{w}_{k+1}, \ldots, \boldsymbol{w}_N]$ and $\boldsymbol{\Sigma} \triangleq [\boldsymbol{B}_1^H \boldsymbol{\Lambda}_1, \boldsymbol{B}_2^H \boldsymbol{\Lambda}_2, \ldots, \boldsymbol{B}_M^H \boldsymbol{\Lambda}_M]$.

Using another set of auxiliary variables $\{\vartheta_n\}_{n \in \mathcal{N}}$, the constraints in (11c) can be reformulated as the intersection of a second-order cone and two positive semidefinite cones as

$$\sqrt{\frac{1+\Gamma_n^{\text{req}}}{\Gamma_n^{\text{req}}}} \boldsymbol{h}_n^H \boldsymbol{w}_n \geq \|[\theta_n, \vartheta_n]\|_F, \forall n \quad (15)$$

$$\begin{bmatrix} \vartheta_n & (P_n^{\text{req}})^{\frac{1}{4}} \\ (P_n^{\text{req}})^{\frac{1}{4}} & \varrho_n \end{bmatrix} \succeq 0 \text{ and } (14), \forall n. \quad (16)$$

As a result, we can check the feasibility of the optimization problem (11) via solving the following problem

$$\text{Find } \{\boldsymbol{w}_n, \rho_n, \varrho_n, \theta_n, \vartheta_n\}_{n \in \mathcal{N}} \text{ and } \{\boldsymbol{\Lambda}_m\}_{m \in \mathcal{M}}$$
$$\text{s.t. } (11a), (11d) - (11e), (13) - (16). \quad (17)$$

We note that the feasible region of (17) is convex; therefore, we can perform the feasibility check via existing toolbox, e.g., CVX [26]. Without loss of generality, we assume that the optimization problem (12) is feasible.

### B. Majorization Procedures

We observe that the objective function in (17) is still non-convex due to the existence of $\ell_0$-norm and the non-convex fronthaul data rate $g_m\left(\{\boldsymbol{w}_n\}_{n \in \mathcal{N}}, \boldsymbol{\Lambda}_m\right), m \in \mathcal{M}$.

*1) Majorization of $\ell_0$-norm:* As proposed in [22], the $\ell_0$-norm, $\left\|\sum_{n=1}^{N} \|\boldsymbol{B}_m \boldsymbol{w}_n\|_F^2\right\|_0$, can be approximated as

$$\left\|\sum_{n=1}^{N} \|\boldsymbol{B}_m \boldsymbol{w}_n\|_F^2\right\|_0 = \lim_{\epsilon \to 0} \frac{\log\left(1 + \sum_{n=1}^{N} \|\boldsymbol{B}_m \boldsymbol{w}_n^{(\tau)}\|_F^2 / \epsilon\right)}{\log(1 + 1/\epsilon)} \quad (18)$$

where $\epsilon$ is a small positive forming factor.

We observe that numerator of (18) is a log-concave function; therefore, we can obtain the upper bound of the numerator of (18) via the first order approximation in the $\tau$-th iteration as

$$\log\left(1 + \epsilon^{-1} \sum_{n=1}^{N} \|\boldsymbol{B}_m \boldsymbol{w}_n\|_F^2\right) \leq \frac{\sum_{n=1}^{N} \|\boldsymbol{B}_m \boldsymbol{w}_n\|_F^2}{\epsilon + \sum_{n=1}^{N} \|\boldsymbol{B}_m \boldsymbol{w}_n^{(\tau)}\|_F^2} - \psi_m^{(\tau)} \quad (19)$$

where

$$\psi_m^{(\tau)} = \frac{\sum_{n=1}^{N} \|\boldsymbol{B}_m \boldsymbol{w}_n^{(\tau)}\|_F^2}{\epsilon + \sum_{n=1}^{N} \|\boldsymbol{B}_m \boldsymbol{w}_n^{(\tau)}\|_F^2}$$
$$- \log\left(1 + \epsilon^{-1} \sum_{n=1}^{N} \|\boldsymbol{B}_m \boldsymbol{w}_n^{(\tau)}\|_F^2\right). \quad (20)$$

*2) Majorization of the Fronthaul Data Rate:* We observe that the fronthaul data rate, $g_m\left(\{\boldsymbol{w}_n\}_{n \in \mathcal{N}}, \boldsymbol{\Lambda}_m\right)$, is the difference of two concave terms. In order to convexify the fronthaul data rate, we surrogate the first concave term via the following procedures

$$\log\det\left(\sum_{n=1}^{N} \boldsymbol{B}_m \boldsymbol{w}_n \boldsymbol{w}_n^H \boldsymbol{B}_m^H + \boldsymbol{\Lambda}_m^2\right)$$
$$\leq \left\|\left(\boldsymbol{U}_m^{(\tau)}\right)^H \boldsymbol{\Lambda}_m\right\|_F^2 + \sum_{n=1}^{N} \boldsymbol{w}_n^H \boldsymbol{B}_m^H \left(\boldsymbol{X}_m^{(\tau)}\right)^{-1} \boldsymbol{B}_m \boldsymbol{w}_n$$
$$+ \log\det \boldsymbol{X}_m^{(\tau)} - i_m \quad (21)$$

where

$$\boldsymbol{X}_m^{(\tau)} \triangleq \sum_{n=1}^{N} \boldsymbol{B}_m \boldsymbol{w}_n^{(\tau)} \left(\boldsymbol{w}_n^{(\tau)}\right)^H \boldsymbol{B}_m^H + \left(\boldsymbol{\Lambda}_m^{(\tau)}\right)^2 \quad (22)$$

and

$$\boldsymbol{U}_m^{(\tau)} \triangleq \boldsymbol{V}_m^{(\tau)} \left(\boldsymbol{D}_m^{(\tau)}\right)^{\frac{1}{2}}. \quad (23)$$

Here, the columns of $\boldsymbol{V}_m^{(\tau)}$ and the diagonal of $\boldsymbol{D}_m^{(\tau)}$ are, respectively, the eigenvectors and eigenvalues of $\left(\boldsymbol{X}_m^{(\tau)}\right)^{-1}$.

## C. Minimization Procedures

Using the surrogate functions in (19) and (21) and introducing a set of auxiliary variables $\{c_m\}_{m\in\mathcal{M}}$, the optimization problem in the minimization procedure is obtained as a convex optimization problem in the $\tau$-th iteration by dropping the constant terms as shown in (26) on top of next page, where $\Delta^{(\tau)}$ and $\delta_m^{(\tau)}$ are, respectively, denoted as

$$\Delta^{(\tau)} \triangleq P^{\text{BBU}} + \sum_{m=1}^{M}\left(P_m^{\text{Sleep}} + \frac{(P_m^{\text{Active}} - P_m^{\text{Sleep}})\psi_m^{(\tau)}}{\log(1+\epsilon^{-1})}\right) \quad (24)$$

$$\delta_m^{(\tau)} \triangleq \frac{P_m^{\text{Active}} - P_m^{\text{Sleep}}}{\log(1+\epsilon^{-1})\left(\epsilon + \sum_{n=1}^{N}\left\|\boldsymbol{B}_m\boldsymbol{w}_n^{(\tau)}\right\|_{\text{F}}^2\right)}. \quad (25)$$

As a result, we propose a GsPSs algorithm whose procedures are summarized in Algorithm 1.

---

**Algorithm 1** GsPSs Algorithm

1: The BBU pool initializes the iteration index $\tau = 0$, the stop threshold $\xi$ and the maximum number of iteration $T_{\max}$
2: The BBU pool obtains the a feasible $\{\boldsymbol{\Lambda}_m^{(0)}\}_{m\in\mathcal{M}}$ and $\{\boldsymbol{w}_n^{(0)}\}_{n\in\mathcal{N}}$ via (17)
3: **repeat**
4: The BBU pool updates $\delta_m^{(\tau)}$ and $\boldsymbol{X}_m^{(\tau)}$ via (20) and (22)
5: With $\{\boldsymbol{\Lambda}_m^{(\tau)}, \delta_m^{(\tau)}, \boldsymbol{X}_m^{(\tau)}\}_{m\in\mathcal{M}}$ and $\{\boldsymbol{w}_n^{(\tau)}\}_{n\in\mathcal{N}}$, the BBU pool solves the optimization problem (26) and obtains the $\{\boldsymbol{\Lambda}_m^{(\tau+1)}\}_{m\in\mathcal{M}}$ and $\{\boldsymbol{w}_n^{(\tau+1)}\}_{n\in\mathcal{N}}$
6: $\tau := \tau + 1$
7: **until** $\frac{\|\boldsymbol{w}_n^\tau - \boldsymbol{w}_n^{\tau-1}\|_{\text{F}}^2}{\|\boldsymbol{w}_n^{\tau-1}\|_{\text{F}}^2} \leq \xi$ and $\frac{\|\boldsymbol{\Lambda}_m^\tau - \boldsymbol{\Lambda}_m^{\tau-1}\|_F^2}{\|\boldsymbol{\Lambda}_m^{\tau-1}\|_F^2} \leq \xi$ or $\tau > T_{\max}$

---

*Proposition 1:* The proposed GsPSs algorithm converges to the sub-optimal solution of the optimization problem (12); therefore, a sub-optimal solution to (9) is obtained.

*Proof:* We denote the optimal solution of the optimization problem (26) in the $\tau$-th iteration as $\{\boldsymbol{w}_n^{(\tau)}, \rho_n^{(\tau)}, \varrho_n^{(\tau)}, \theta_n^{(\tau)}, \vartheta_n^{(\tau)}\}_{n\in\mathcal{N}}$ and $\{\boldsymbol{\Lambda}_m^{(\tau)}, c_m^{(\tau)}\}_{m\in\mathcal{M}}$. The optimal solution $\{\boldsymbol{w}_n^{(\tau)}, \rho_n^{(\tau)}, \varrho_n^{(\tau)}, \theta_n^{(\tau)}, \vartheta_n^{(\tau)}\}_{n\in\mathcal{N}}$ and $\{\boldsymbol{\Lambda}_m^{(\tau)}, c_m^{(\tau)}\}_{m\in\mathcal{M}}$ lie in the feasible region of the $(\tau+1)$-th iteration. As a result, the objective value decreases after each iteration. The objective value is lower bounded due to the constraints in (11b) and (11c). Hence, the proposed Algorithm 1 converges. With the similar arguments in [27], we can prove the convergent solution $\{\boldsymbol{w}_n^{(\infty)}, \rho_n^{(\infty)}, \varrho_n^{(\infty)}, \theta_n^{(\infty)}, \vartheta_n^{(\infty)}\}_{n\in\mathcal{N}}$ and $\{\boldsymbol{\Lambda}_m^{(\infty)}, c_m^{(\infty)}\}_{m\in\mathcal{M}}$ satisfies the Karush-Kuhn-Tucker conditions of (12). Based on [11, Proposition 3.1], the Algorithm 1 converges to the sub-optimal solution of (9). □

## IV. NUMERICAL RESULTS

In this section, we present the numerical results to demonstrate the effectiveness of the proposed GsPSs algorithm. Unless otherwise specified, the simulation parameters are specified in Table I.

Figure 1 shows the convergence behavior with different values of the forming factor $\epsilon$. We observe that the system power consumption decreases monotonically and converges to the suboptimal objective values after a certain number of iterations (less than 10 iterations) under different forming factors, i.e., $\epsilon = 10^{-1}$, $\epsilon = 10^{-2}$, $\epsilon = 10^{-3}$ and $\epsilon = 10^{-4}$.

TABLE I
SIMULATION PARAMETERS SETTING

| Parameters | Values |
|---|---|
| Number of RRHs, $M$ | 5 |
| Number of UEs, $N$ | 3 |
| Number of Antenna per RRH | 3 |
| Gaussian noise power, $\sigma_{n,\text{c}}^2$ | $-30$ dBm |
| Signal processing noise power, $\sigma_{n,\text{p}}^2$ | $-30$ dBm |
| Circuit power consumption | $P_m^{\text{SP}} = 27$ dBm<br>$P_m^{\text{Sleep}} = 13$ dBm<br>$P^{\text{BBU}} = 23$ dBm |
| Maximum TX power of RRH, $P_m^{\max}$ | 20 dBm |
| Minimum required input power, $P_n^{\text{req}}$ | 0 dBm |
| SINR requirement, $\Gamma_n^{\text{req}}$ | 10 dB |
| Fronthaul power cost factor, $\zeta_m$ | 7 dBm/(nat/sec/Hz)$^2$ |
| Amplifier efficiency, $\phi_m$ | 0.8 |
| Pathloss factor, $\chi$ | 2.3 |

We also observe that a smaller value of $\epsilon$ results in a smaller objective value of (26). This is due to the fact that the surrogate function can perform a better approximation of the $\ell_0$-norm with a smaller value of $\epsilon$. Thus, we use $\epsilon = 10^{-4}$ in the following numerical results.

Figure 2 illustrates the performance variation against the maximum transmission power of each RRH. In Fig. 2(a), we observe that the total power consumption decreases when $P_m^{\max}$ is smaller than 24 dBm. The reasons can be explained as follows. The larger maximum transmission power $P_m^{\max}$ of each RRH leads to a smaller number of active RRHs as shown in Fig. 2(b). The smaller number of active RRHs indicates a smaller amount of circuit power consumption. Therefore, the total power consumption is decreased. As the maximum transmission power $P_m^{\max}$ continues to increase, the total power consumption saturates. This observation can be justified as follows. When the number of active RRHs can satisfy the data rate requirements of the IDMs and the input power demands of the EHMs, switching off more RRHs causes excessive consumed power for those active RRHs.

## V. CONCLUSIONS

We investigated the power consumption minimization problem in the SWIPT C-RANs. In this paper, the power consumption is defined as the summation of the transmission power, the circuit power and the fronthaul power consumption. We proposed a more practical nonlinear model for the fronthaul links power consumption. Based on the nonlinear model and the MM framework, an algorithm based on group sparse precoding and signal splitting was proposed to obtain the suboptimal solution of the precoding coefficients, the set of active RRHs and the signal splitting ratios. Numerical results show that the proposed algorithm converges after a certain number of iterations. Our results unveil that increasing the power budget of each RRH does not necessarily reduce power consumption.

## ACKNOWLEDGE

This work was supported in part by the University of British Columbia Four-Year Fellowship and in part by a Canadian National Sciences and Engineering Research Council Strategic Partnership Grant.

$$\min_{\substack{\{\boldsymbol{w}_n,\rho_n,\varrho_n,\theta_n,\vartheta_n\}_{n\in\mathcal{N}} \\ \{\boldsymbol{\Lambda}_m,c_m\}_{m\in\mathcal{M}}}} \sum_{n=1}^{N}\sum_{m=1}^{M}\left(\frac{1}{\phi_m}+\delta_m^{(\tau)}\right)\|\boldsymbol{B}_m\boldsymbol{w}_n\|_{\mathrm{F}}^2 + \sum_{m=1}^{M}\frac{1}{\phi_m}\|\boldsymbol{\Lambda}_m\|_{\mathrm{F}}^2 + \sum_{m=1}^{M}\zeta_m c_m^2 + \Delta^{(\tau)}$$

$$\text{s.t.} \sum_{n=1}^{N}\boldsymbol{w}_n^{\mathrm{H}}\boldsymbol{B}_m^{\mathrm{H}}\left(\boldsymbol{X}_m^{(\tau)}\right)^{-1}\boldsymbol{B}_m\boldsymbol{w}_n + \left\|\left(\boldsymbol{U}_m^{(\tau)}\right)^{\mathrm{H}}\boldsymbol{\Lambda}_m\right\|_{\mathrm{F}}^2 - 2\log\det\boldsymbol{\Lambda}_m \leq c_m + i_m - \log\det\boldsymbol{X}_m^{(\tau)}, \forall m$$

$$(11\mathrm{a}), (11\mathrm{d})-(11\mathrm{e}), (13)-(16) \tag{26}$$

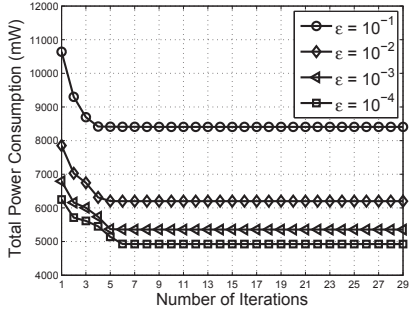

Fig. 1. The convergence behavior versus the forming factor $\epsilon$.

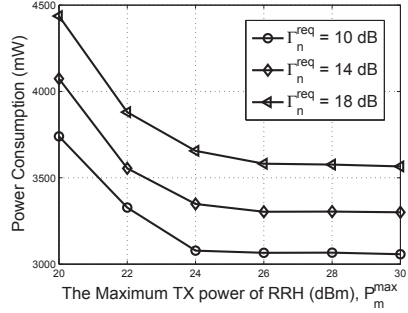

(a) Total transmission power v.s. $P_m^{\max}$.

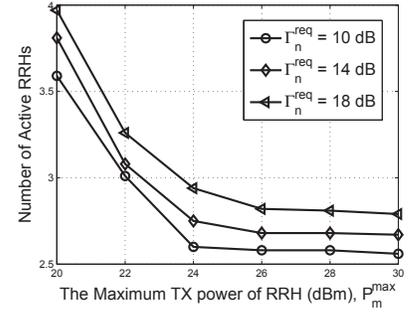

(b) The number of active RRHs v.s. $P_m^{\max}$.

Fig. 2. The performance variation against the maximum transmission power of each RRH, $P_m^{\max}$.

## REFERENCES

[1] B. Kellogg, V. Talla, J. R. Smith, and S. Gollakot, "PASSIVE WI-FI: Bringing low power to Wi-Fi transmissions," *GetMobile: Mobile Comput. and Commun.*, vol. 20, no. 3, pp. 38–41, Jan. 2017.

[2] X. Lu, P. Wang, D. Niyato, D. I. Kim, and Z. Han, "Wireless networks with RF energy harvesting: A contemporary survey," *IEEE Commun. Surveys Tuts.*, vol. 17, no. 2, pp. 757–789, Second quarter 2015.

[3] X. Zhou, R. Zhang, and C. K. Ho, "Wireless information and power transfer: Architecture design and rate-energy tradeoff," *IEEE Trans. Commun.*, vol. 61, no. 11, pp. 4754–4767, Nov. 2013.

[4] A. Checko, H. L. Christiansen, Y. Yan, L. Scolari, G. Kardaras, M. S. Berger, and L. Dittmann, "Cloud RAN for mobile networks – a technology overview," *IEEE Commun. Surveys Tuts.*, vol. 17, no. 1, pp. 405–426, First quarter 2015.

[5] H. Zhang, Y. Dong, J. Cheng, M. J. Hossain, and V. C. M. Leung, "Fronthauling for 5G LTE-U ultra dense cloud small cell networks," *IEEE Wireless Commun.*, vol. 23, no. 6, pp. 48–53, Dec. 2016.

[6] M. Peng, Y. Li, Z. Zhao, and C. Wang, "System architecture and key technologies for 5G heterogeneous cloud radio access networks," *IEEE Netw.*, vol. 29, no. 2, pp. 6–14, Mar. 2015.

[7] Z. Ding, C. Zhong, D. W. K. Ng, M. Peng, H. A. Suraweera, R. Schober, and H. V. Poor, "Application of smart antenna technologies in simultaneous wireless information and power transfer," *IEEE Commun. Mag.*, vol. 53, no. 4, pp. 86–93, Apr. 2015.

[8] L. Liu, R. Zhang, and K. C. Chua, "Wireless information and power transfer: A dynamic power splitting approach," *IEEE Trans. Commun.*, vol. 61, no. 9, pp. 3990–4001, Sept. 2013.

[9] G. Pan, C. Tang, T. Li, and Y. Chen, "Secrecy performance analysis for SIMO simultaneous wireless information and power transfer systems," *IEEE Trans. Commun.*, vol. 63, no. 9, pp. 3423–3433, Sept. 2015.

[10] L. Chen, F. R. Yu, H. Ji, B. Rong, X. Li, and V. C. M. Leung, "Green full-duplex self-backhaul and energy harvesting small cell networks with massive MIMO," *IEEE J. Sel. Areas Commun.*, vol. 34, no. 12, pp. 3709–3724, Dec. 2016.

[11] Q. Shi, W. Xu, T. H. Chang, Y. Wang, and E. Song, "Joint beamforming and power splitting for MISO interference channel with SWIPT: An SOCP relaxation and decentralized algorithm," *IEEE Trans. Signal Process.*, vol. 62, no. 23, pp. 6194–6208, Dec. 2014.

[12] Y. Dong, H. Zhang, M. J. Hossain, J. Cheng, and V. C. M. Leung, "Energy efficient resource allocation for OFDMA full duplex distributed antenna systems with energy recycling," in *Proc. of IEEE Globecom*, Dec. 2015, pp. 1–6.

[13] D. W. K. Ng, Y. Wu, and R. Schober, "Power efficient resource allocation for full-duplex radio distributed antenna networks," *IEEE Trans. Wireless Commun.*, vol. 15, no. 4, pp. 2896–2911, Apr. 2016.

[14] H. Zhang, J. Du, J. Cheng, and V. C. M. Leung, "Resource allocation in SWIPT enabled heterogeneous cloud small cell networks with incomplete CSI," in *Proc. of IEEE Globecom*, Dec. 2016, pp. 1–5.

[15] Y. Gu and S. Aïssa, "RF-based energy harvesting in decode-and-forward relaying systems: Ergodic and outage capacities," *IEEE Trans. Wireless Commun.*, vol. 14, no. 11, pp. 6425–6434, Nov. 2015.

[16] E. Boshkovska, D. W. K. Ng, N. Zlatanov, and R. Schober, "Practical non-linear energy harvesting model and resource allocation for SWIPT systems," *IEEE Commun. Lett.*, vol. 19, no. 12, pp. 2082–2085, Dec. 2015.

[17] Y. Dong, M. J. Hossain, and J. Cheng, "Performance of wireless powered amplify and forward relaying over Nakagami-$m$ fading channels with nonlinear energy harvester," *IEEE Commun. Lett.*, vol. 20, no. 4, pp. 672–675, Apr. 2016.

[18] A. E. Shafie, D. Niyato, and N. Al-Dhahir, "Security of an ordered-based distributive jamming scheme," *IEEE Commun. Lett.*, vol. 21, no. 1, pp. 72–75, Jan. 2017.

[19] D. W. K. Ng and R. Schober, "Secure and green SWIPT in distributed antenna networks with limited backhaul capacity," *IEEE Trans. Wireless Commun.*, vol. 14, no. 9, pp. 5082–5097, Sept. 2015.

[20] B. Dai and W. Yu, "Energy efficiency of downlink transmission strategies for cloud radio access networks," *IEEE J. Sel. Areas Commun.*, vol. 34, no. 4, pp. 1037–1050, Apr. 2016.

[21] D. L. Donoho, "Compressed sensing," *IEEE Trans. Inf. Theory*, vol. 52, no. 4, pp. 1289–1306, Apr. 2006.

[22] B. K. Sriperumbudur, D. A. Torres, and G. R. G. Lanckriet, "A majorization-minimization approach to the sparse generalized eigenvalue problem," *Mach. Learn.*, vol. 85, no. 1, pp. 3–39, Oct. 2011.

[23] T. M. Cover and J. A. Thomas, *Elements of Information Theory*. New York, NY, USA: John Wiley & Sons, 2012.

[24] Y. Dong, X. Ge, M. J. Hossain, J. Cheng, and V. C. M. Leung, "Proportional fairness based beamforming and signal splitting for MISO-SWIPT systems," *IEEE Commun. Lett.*, to be published.

[25] A. Attar, H. Li, and V. C. M. Leung, "Green last mile: How fiber-connected massively distributed antenna systems can save energy," *IEEE Wireless Commun.*, vol. 18, no. 5, pp. 66–74, Oct. 2011.

[26] M. Grant and S. P. Boyd, "CVX: Matlab software for disciplined convex programming, version 2.1," http://cvxr.com/cvx, Mar. 2014.

[27] A. Beck, A. Ben-Tal, and L. Tetruashvili, "A sequential parametric convex approximation method with applications to nonconvex truss topology design problems," *J. Glob. Optim.*, vol. 47, no. 1, pp. 29–51, 2009.